\documentclass[11pt]{amsart}

\usepackage{amsmath,graphics}
\usepackage{amsfonts,amssymb}

\begin{document}

\title{An exotic proposal on Cooper pairing in high-$T_c$ supercondutors}
\author{Yanghyun Byun}

\address{Department of Mathematics, Hanyang University
(yhbyun@hanyang.ac.kr)}

\begin{abstract}
We propose that a Cooper pair in a high-$T_c$ superconductor might be in fact a bound system $\rm eex\bar x$
of two electrons, a particle $\rm x$  and its antiparticle. We assume $\rm eex\bar x$ is more massive
than two free electrons.
We observe that $\rm eex\bar x$ should be stable in a solid if it is in a low enough state in the solid.
A solid which admits such a low state for $\rm eex\bar x$ has some properties which seem closely related to the behavior
of HTS materials. The interaction of the $\rm x$-particle which binds the four particles into
$\rm eex\bar x$ has been discussed. We also discuss the scale of the mass excess of $\rm eex\bar x$ relative to two free electrons.
An experiment designed to detect a bound system of two electrons with a mass excess has been proposed.
\end{abstract}

\maketitle

\section{The proposal}
After more than a quarter century since its discovery\cite{1} there are many theories on high-$T_c$ superconductivity.
However any of them has not yet earned recognition by a majority of physicists.
The recent quarter century can be viewed as a more frustrating experience than the previous half century
in which physicists struggled for an understanding of conventional superconductivity until the emergence of BCS theory\cite{2}.
Indeed, one may say that the total amount and the quality of the labor put into the current problem
have exceeded those put into the previous one.

We tend to take it for granted that any theory of high-$T_c$ superconductivity should be built
on the known first principles, that is, on the Standard Model,
even if only a small portion of the semi-complete theory is relevant.
Any attempt not based on the SM may be regarded as exotic.
However the author believes the current status of affair has made considering
an extraordinary possibility appear making more sense than sometime before.
In this paper we investigate closely an exotic possibility concerning the Cooper
pair in high-$T_c$ superconductors.

In cuprates the Cooper pair size is estimated to be about $\rm 1-2\, nm$\cite{3} and in the more recent iron pnictides,
about $\rm 3\, nm$\cite{4}. They are smaller by 2 orders of magnitude than the Cooper pair size in conventional superconductors.
On the other hand a calculation has shown that a stable positronium ion $\rm ee\bar e$ exists and its size is about $\rm 0.5\, nm$\cite{5}.
A stable bound system  $\rm eee\bar e$, if one exists, will be more relevant
to the Cooper pair and its size must be considerably larger than $\rm 0.5\, nm$.
Thus one may view the size of a Cooper in an HTS as comparable
to the scale of an atomic system consisting of some light particles.
Furthermore some physicists suspect that the Cooper pairs in an HTS are preformed
far above the superconducting transition temperature\cite{6}.
This may mean that the formation of Cooper pairs is irrelevant to lowering the temperature. We propose:

 \begin{quote}
 {\bf Proposal:} A Cooper pair in an HTS is in fact a bound system  $\rm eex \bar x$ of two electrons, a particle $\rm x$
 and its anti-particle.
 \end{quote}

The binding mechanism is discussed in \S 3 below. We consider the proposal together with the following:

\begin{quote}
{\bf Assumption 1:} The bound system $\rm eex \bar x$ is more massive than two free electrons.
\end{quote}
Otherwise $\rm eex \bar x$ could have existed stably in free space.
The experiment proposed in \S 5 below,  if it works, can measure the excess of mass.
We propose that the excess is greater than $\rm 4\, eV$ and less than $\rm 200\, eV$ in \S 4.

An urgent problem is how a system containing a particle and its anti-particle can be stable.
To provide a mechanism, let $E_0 < 0$ denote 2 times the energy of the lowest unoccupied electron state of a solid.
On the other hand we denote by $E_x$ the energy of the lowest state of $\rm eex \bar x$ in the solid plus the mass energy gain
of $\rm eex \bar x$ relative to two free electrons.
Then the system $\rm eex \bar x$ in its lowest state will be stable in the solid if the inequality $E_x <E_0 $ holds.
This mechanism can be compared to the one by which a neutron is stable in a nucleus.
Therefore we assume:

\begin{quote}
{\bf Assumption 2:} The inequality  $E_x <E_0 <0$ holds in an HTS.
\end{quote}

An important merit of the $\rm eex \bar x$-hypothesis is that $\rm x$ and $\rm \bar x$,
being a particle and its antiparticle, may come into existence without violating conservation laws
as follows:
\begin{center}
{\small (a virtual boson which is its own antiparticle)
  $\rightarrow$ (a virtual pair  of $\rm x$ and $\rm \bar x$ ).}
\end{center}
\noindent And they may become real and stable by being caught in $\rm eex \bar x$ which is in a low enough state
in a solid, even if we may not yet describe the process in details.
The virtual boson may as well be the photon. If it is the case, $\rm x$ and $\rm \bar x$
should be electrically charged.
However we argue in \S 3 below that $\rm x$ and $\rm \bar x$ must be able
to interact with the electron
by some other means than the Coulomb force.
Therefore we may not regard the photon as the only option for the boson above.

Consider the HTS, $\rm YBa_2Cu_3O_{7-\delta}$, and note that the innermost state of an electron
in the heaviest atom $\rm {}^{52}Ba$ has an energy about $\rm -35\ keV$. Apparently there would have been no chance
that the inequality $E_x <E_0 $ could be satisfied if $\rm x$ were the electron. The same applies to any charged leptons.
In fact it is clear that $\rm x$ cannot be any of the particles listed in the Standard Model.
Thus the $\rm eex \bar x$-hypothesis is a proposal of a new particle for the purpose of explaining
a phenomenon below room temperature. On the other hand the SM is a tight network of a myriad of precise measurements
interwoven by quantum field theory. In the theory a fundamental particle may influence events
by participating as virtual particles. Therefore any hypothetical new particle should be
such that it is not only difficult to be observed directly but also its existence does not
affect significantly
the calculations of SM which are supported by precise measurements.

If we assume that an $\rm x$-particle is always confined in the bound system $\rm x \bar x$,
in a similar way as a quark is confined, it might explain how it could have been unnoticed so far.
It is also possible that a free $\rm x$-particle is too massive to be created in modern accelerators.
If a free $\rm x \bar x$-particle is massive enough, it will explain how the particle could have been hidden from observations.
Even if the mass of $\rm x \bar x$ is in an accessible range, it is not clear to the author
whether its presence could have been easily noticed if $\rm x \bar x$ may decay into only two or more photons.
When one considers the effects of virtual  $\rm x$'s, it might be necessary
to assign a large enough value for the mass of virtual $\rm x$
or for the mass of the virtual boson which gives rise to the pair $\rm x$ and $\rm \bar x$
so that the calculations of SM up to now can be kept without significant changes.
It is also possible that the interaction of $\rm x$ has a small coupling constant or
that the  particle is electrically neutral and its interaction happens to be irrelevant
to the physical values at which the calculations are aimed.
It seems that there can be more than one scenario by which the $\rm x$-particle could have been unnoticed
by modern physics.

There are immediate consequences regarding the properties of a solid that can be derived
from the assumption that $\rm eex \bar x$'s are present in the solid. They seem closely related to the behaviors
of an HTS and are discussed in \S 2.
The interaction of $\rm x$-particle is discussed  in \S 3 together with the mechanism which binds the four particles
into $\rm eex \bar x$. A reader may skip \S 3 at first reading.
We discuss the scale of the mass excess of $\rm eex \bar x$ in \S 4.
An experiment designed to detect a bound state of two electrons
with a mass excess has been proposed in \S 5.

\section{Relevance to superconductivity}
Assume the inequality $E_x <E_0$ holds in a solid and the solid is at absolute zero temperature.
Then there can be no electron in a state whose energy is greater than $\frac{1}{2} E_x$:
There is no such thing as a filled state for $\rm eex \bar x$ which is a boson.
And any two electrons in states with energies above $\frac{1}{2} E_x$ should form $\rm eex \bar x$ and fall
into a sate with apparent energy $\frac{1}{2} E_x$.
Therefore to an observer who does not suspect the presence of $\rm eex \bar x$,
the electrons in $\rm eex \bar x$ will appear to be in the highest energy state.

Now assume that $\rm eex \bar x$'s exist in a solid and $\frac{1}{2} E_x$ is
the energy of a state in a partially filled band. Then we must have the equality  $E_x = E_0$ instead of the inequality
$E_x <E_0$: Since the band is partially filled there are electrons in states with energies infinitesimally close
to $\frac{1}{2} E_0$.
If the inequality holds it is energetically advantageous for electrons in those states to form $\rm eex \bar x$
so that they may apparently be in a state with energy $\frac{1}{2} E_x$. The equality $E_x = E_0$ implies that $\rm eex \bar x$'s
are not stable but in an equilibrium with the unbound electrons in states with energy near $\frac{1}{2} E_x$.

Thus only when all the bands which contain states with energy lower than $\frac{1}{2} E_x$ is filled,
the inequality $E_x <E_0$ may hold. In particular only in this case an $\rm eex \bar x$ may be stable.
Furthermore we note that this means $\frac{1}{2} E_x$ should lie in the energy gap of the band structure
below which all electron states are occupied and above which no state is occupied.

Now let us assume that a necessary condition for high-$T_c$ superconductivity is that an $\rm eex \bar x$ is stable
since it is less likely that $\rm eex \bar x$'s may carry a supercurrent if some of them are consistently decaying
into unbound pairs of electrons. This means that the inequality $E_x <E_0$ holds instead of the equality $E_x = E_0$ in an HTS.
On the other hand mobility must be essential for $\rm eex \bar x$'s to carry any electric current.
Note that the energy of the lowest state of $\rm eex \bar x$ in the solid is lower than $E_x$
by the mass energy gain of $\rm eex \bar x$ relative to two free electrons.
Therefore the condition that $\rm eex \bar x$'s should be mobile might be one that cannot be easily met.
Even if $E_x <E_0$ holds and $\rm eex \bar x$'s are mobile in some material, something like a superfluid phase of $\rm eex \bar x$'s
might be necessary as well for the onset of superconductivity\cite{7}.

In the scenario above the formation of $\rm eex \bar x$ has nothing to do with lowering the temperature.
Indeed some physicists suspect preformed Cooper pairs in HTS materials far above the transition temperature\cite{6}.
However a low temperature is needed to keep $\rm eex \bar x$'s stable: The temperature is limited by the gaps from
$\frac{1}{2} E_x$ to the top of the filled bands and to the bottom of the unfilled bands.
It is not clear
whether these gaps are directly related to the so-called superconducting energy gap of an HTS.
Indeed, it might be the case that the states of $\rm eex \bar x$ in the solid form a band structure or
that there are more than one states for the bound system $\rm eex \bar x$ itself.
A low enough temperature might be necessary also for a superfluid phase.

Again assume the zero temperature. In case the equality $E_x = E_0$ holds,
the electrons at the apparent Fermi surface are in fact electrons in the bound system $\rm eex \bar x$.
In case the inequality $E_x <E_0$ holds the apparent Fermi band has zero width consisting only of $\rm eex \bar x$'s.
It is not clear to the author whether these statements can be reconciled with the measurements made on HTS's.
However it seems worth a note that the Fermi energy of an HTS is known to be extremely small\cite{8}.
Now the following assumption seems not only necessary but also reasonable:
\begin{quote}
{\bf Assumption 3:} As the density of $\rm eex \bar x$ per unit volume increases $E_x$ rises.
\end{quote}
\noindent At fine tuning this assumption may account for the low carrier density\cite{9}
of an HTS by limiting the density of $\rm eex \bar x$.

On the other hand we note that most of the arguments above are implicitly based on the assumption
that $E_x$ remains constant. Now they may not be valid anymore: Let $E_x(\rho)$ denote $E_x$
when the density of $\rm eex \bar x$ per unit volume is $\rho$. Also let $V$ be the volume of the solid.
Then depending on the type of the function $E_x$ of $\rho$, the total energy $E_x(\rho)\rho V$
may have the minimum at some density $\rho_0$ and it might happen $E_x(\rho_0) <E_f$, where $E_f$ denotes 2 times
the highest energy of the filled electron states. This puts in doubt all the conclusions above regarding a solid
at the absolute zero temperature. Nevertheless they will remain valid if the type of the function $E_x$
of $\rho$ is appropriate. More importantly the author believes the effects of presence
of $\rm eex \bar x$ in a solid are more efficiently illustrated with the simple assumption that $E_x$ is constant.

For an HTS we are still preserving the inequality $E_x <E_0$ of Assumption 2.
Note that HTS's are often doped systems. The doping might be a process by which a material supporting the inequality
$E_x <E_0$ is obtained from a material in which $E_x \geq E_0$ holds. We provide a scenario
by which doping induces such a transformation as follows: Consider a material which needs electron doping to superconduct.
Assume for simplicity that the doping does not change the band structure significantly
except that it provides extra electrons to the Fermi band. We may assume that the material supports the equality $E_x = E_0$
from the beginning or that it has transformed, by some dose of doping, into such a material.
Then $\frac{1}{2} E_x$ is the energy of the Fermi surface as observed in the above.
If we keep adding electrons to the Fermi band, $E_x$ should eventually be elevated and the Fermi band should be filled.
For if $E_x$ is left constant all the added electrons should pairwise form $\rm eex \bar x$'s
since all electron states below $\frac{1}{2} E_x$ is filled and there cannot be any electron in a state above $\frac{1}{2} E_x$,
which will increase the density of $\rm eex \bar x$, contradicting Assumption 3. Now we need a gap above the Fermi band.
Then all the electrons added right after the Fermi band is filled will pairwise form $\rm eex \bar x$'s
and $\frac{1}{2} E_x$ will enter the gap.

Even if $\rm eex \bar x$ turns out to be real, a condensed matter physics incorporating
the existence of $\rm eex \bar x$ should be developed
before a full explanation of high-$T_c$ superconductivity.
And such an accomplishment may come only
after proper understanding of the system $\rm eex \bar x$ itself,
its interaction with the lattice and the interaction between themselves.
The author admits that the
$\rm eex \bar x$-hypothesis does not simplify the problem.

\section{On the interaction}
We begin by noting that the particles $\rm x$ and $\rm \bar x$ should be spatially separated
enough for them to interact with other particles. However it must be that $\rm x$ and $\rm \bar x$
are still in a bound state with each other since it is unlikely that free $\rm x$-particles
could have been hiding unnoticed from the eyes of modern physics. One may imagine
that the $\rm x$-particles are confined in the bound system $\rm x\bar x$
within the distance of a Cooper pair size, that is, within $\rm 1{\text -}3\, nm$, in a similar way as quarks
are confined in a hardron within the distance of about $\rm 2 \times 10^{-6}\, nm$.

Let us suppose that the only interaction of $\rm x$
with the electrons and the nuclei is electric. We also assume that $0<q_x \leq 2e$,
where $q_x$  is the electric charge of $\rm x$
which we choose to be positive. Otherwise $\rm e_k x \bar x$ with $\rm k \geq 3$
seems more likely than $\rm eex\bar x$.
We show in the below that the combination of these two assumptions is unrealistic.

Firstly we consider the case when $q_x =e$. Then the four particles may bind together
to form $\rm eex\bar x$ as follows: It is known that, if $\rm y$  is any particle with the electric charge $e$,
a stable bound state $\rm eey$ exists regardless of the mass of $\rm y$,
which is illustrated by the existence of the hydrogen ion $\rm H_2^{+}$, the negative hydrogen ion $\rm H^{-}$
and the positronium ion $\rm ee\bar e$\cite{10}.
If we assume that $\rm x$ can move about freely within the distance
of $\rm 1{\text -}3\, nm$ from $\rm \bar x$, a stable bound system $\rm eex$ may form,
even if a worry here is that the size must be larger than that of $\rm ee\bar e$
and might be comparable or larger than $\rm 3\, nm$ if the mass of $\rm x$ is much smaller than that of the electron.
Since $\rm \bar x$ is still bound to $\rm x$, by some other interaction than the Coulomb force,
we have a stable bound system $\rm eex {\text -} \bar x$.

However there seems a decisive reason that the picture above cannot be realistic:
The absolute value of the potential energy of $\rm x$ in $\rm eex$ due to the two electrons
must be much smaller than that of the proton in a hydrogen ion $\rm H^{-}$.
Let us take as a lower bound of the potential energy, rather generously,
two times the potential energy of the proton in a hydrogen atom which is about $\rm -54\, eV$.
Since as a whole $\rm x \bar x$ is bound to $\rm ee$ by the Coulomb force, the kinetic energy of $\rm x \bar x$
should be less than $\rm 54\, eV$. Also $\rm x\bar x$ can be regarded as confined
within a region with diameter $\lambda =\rm 3\, nm$ around $\rm ee$.
A particle confined in a 1 dimensional interval of length $\lambda$
with kinetic energy $E_k$ has the mass energy greater than $E_m =\frac{1}{2E_k}(\frac{hc}{\lambda})^2$.
Assuming the confinement is within a 3 dimensional cube and the kinetic energy $E_k$ is shared equally
by the three components, we have $E_m = \frac{3}{2E_k}(\frac{hc}{\lambda})^2$.
In our case $E_m$ is about $\rm 4.7\, keV$. Thus if $m_{x\bar x}$ denotes the mass of $\rm x\bar x$
we have the inequality:
$$ m_{x\bar x}c^2 > {\rm 4.7\, keV}\, .$$

On the other hand, even if we assume that $\bar x$ is in the lowest state of the heaviest atom,
the energy can account for only a small portion of the mass energy of $\rm x\bar x$:
We consider the HTS, $\rm YBa_2Cu_3O_{7-\delta}$, and note that the energy of the innermost electrons
in the heaviest atom $\rm {}^{52}Ba$ is roughly $E_i \approx {\rm 51^2 (-13.6\, eV) \approx -35 \, keV}$.
The energy of the innermost state of $\bar x$ is given by $\frac{m_x}{m_e}E_i =m_x \frac{E_i}{m_e}$.
Since $\frac{E_i}{m_e}= -0.07 c^2$  and $m_x \leq m_{x\bar x}$, most of the mass energy $m_{x\bar x}c^2$
should be absorbed by the interaction of $\rm eex$  with the lattice.
This requires that the potential energy due to the subsystem $\rm eex$ in the solid should be below $\rm -4.4\, keV$.
However $\rm eex$ under current assumptions is such a feeble system that it cannot withstand
such a strong field as demanded by the potential. Thus it is unrealistic to assume that the only interaction of $\rm x$
with the electrons and the nuclei is electric under the condition $q_x =e$.

Secondly we consider the case when $q_x =2e$.
Then a helium-like system $\rm eex$ can be considered and subsequently a bound system $\rm eex{\text -}\bar x$
may exist in a similar way as in the above. The potential energy of $\rm x$ in $\rm eex$ due to the two electrons
must be much smaller in its absolute value than that of the nucleus of the helium atom which is about $\rm -180\, eV$.
We take $\rm 180\, eV$ as an upper bound for the kinetic energy of $\rm x\bar x$ in $\rm eex \bar x$.
Now the same argument as in the above can be applied here to show that $m_{x\bar x}c^2 > {\rm 1.4\, keV}$.
Again most of the mass energy $m_{x\bar x}c^2$ should be absorbed by the interaction of $\rm eex$ with the lattice,
say, of the $\rm YBa_2Cu_3O_{7-\delta}$ superconductor.
However since $\rm eex$ is electrically neutral it may interact with the lattice
only by a Van der Waals force and it is very unlikely that
it may be in a state with energy lower than $\rm -1.3\, keV$.

Similar arguments applies to all the cases $0< q_x \leq 2e$:
There will be no bound system $\rm eex\bar x$ at all if $q_x$ is too small.
Otherwise a bound system $\rm eex{\text -}\bar x$ can be considered.
However the electric potential of $\rm x$ in $\rm eex$ due to the two electrons is so small that, for $\rm x\bar x$
to be bound to the two electrons within the distance of $\rm 3\, nm$
the mass of $x\bar x$ must be greater than $\rm 1.4\, keV$.
On the other hand since the interaction of $\rm \bar x$ with the lattice of an HTS, say, $\rm YBa_2Cu_3O_{7-\delta}$,
may account for only small portion of its own mass energy,
most of the mass energy of $\rm x\bar x$ should be absorbed by the electric interaction of $\rm eex$ with the lattice.
However it is unlikely that $\rm eex$ may have a potential lower than $\rm -1.3\, keV$ not being torn apart.
We conclude:
\begin{quote}
It is impossible that the Coulomb force is the only interaction
of $\rm x$ with the electrons
and the nuclei under the condition $0<q_x \leq 2e$.
\end{quote}

We propose that the inequality $0 \leq q_x \leq 2e$ should be kept.
If we regard $\rm x$ as a fundamental particle, this looks quite reasonable
since there is no known fundamental particle with electric charge larger than $e$.
Then the conclusion above implies:
\begin{quote}
The particle $\rm x$ must interact with the electrons
or with the nuclei by some other means than the Coulomb force.
\end{quote}

The conclusion above may mean an electron or a nucleus has a new kind of charge.
Now the $\rm eex\bar x$-hypothesis appears even more radical and alarming
than when it is understood simply as hypothesizing a new particle.

We speculate on the nature of the interaction of $\rm x$ as follows:
First of all there is little possibility of a significant long range non-electric interaction
of $\rm x$ with electrons or with the nuclei, where `long range' refers to a distance
farther than the typical size of an atom in a solid. For if there was a long range interaction of $\rm x$
with the lattice causing more than a few $\rm eV$ energy difference to the states of $\rm eex\bar x$,
we may expect the same kind of interaction among electrons or among nuclei and the interaction should have been noticed
by material science. Therefore the interaction of $\rm x$ with the SM particles is most likely of a short range.
On the other hand if a short range interaction with the nuclei were responsible
for a low enough potential of $\rm eex\bar x$, it would have been impossible
that an $\rm eex\bar x$ may be mobile in the solid.
However apparently there are mobile states of $\rm eex\bar x$ in an HTS.
Thus the only remaining option is that $\rm x$ interacts with the electron by a short range force.
We may expect that the short range interaction is present also between electrons.
Since electrons are densely packed near a heavy nucleus and no effect of exotic interaction
between the electrons has been noticed, the interaction is most likely of much shorter range
than the typical distance between electrons in the innermost shells of heavy atoms.
On the other hand we might have to believe that the interaction is present also
in the electron-positron scattering in which the electron and positron
approach each other within very short distance.
However it is not clear what kind of anomaly should be expected in the experiments
due to the presence of such an interaction.
Now we summarize the speculations so far as follows:
\begin{quote}
{\bf Assumption 4:} The particle $\rm x$  interacts with the electron
by a short range force to form $\rm eex\bar x$ while it may interact by any other non-electric force
neither with electrons nor with nuclei in any significant way.
\end{quote}

Now we have three choices, $\rm ex{\text -}e\bar x$, $\rm eex{\text -}\bar x$ and $\rm x{\text -}ee\bar x$,
for the structure of $\rm eex\bar x$.
We may assume all these candidates are sturdy unlike the feeble system $\rm eex{\text -}\bar x$ in the above
in which the subsystem $\rm eex$ was maintained only by the Coulomb force.
On the other hand the experiment proposed in \S 5 below is only such that may test whether there is
a bound system of two electrons with a mass excess and it is not desirable for us to speculate further
on the interaction of $\rm x$ at this point.
In the next section we try to have some idea on the scale of the mass excess of $\rm eex\bar x$.
In the argument the actual structure of $\rm eex\bar x$ will not be very important.

\section{On the mass excess}

Recall we denote by $q_x$ the electric charge of $\rm x$.
In this section we assume $0 \leq q_x \leq e$.
On the other hand Assumption 4 above implies:
\begin{quote}
The bound system $\rm eex\bar x$ interacts
with the lattice only by the electric force.
\end{quote}
Furthermore considering the previous section we assume: (1)There are three options,
$\rm ex{\text -}e\bar x$, $\rm eex{\text -}\bar x$ and $\rm x{\text -}ee\bar x$, for the structure of $\rm eex\bar x$,
which we leave undetermined in this paper. (2)Whatever the actual structure is,
a subsystem of $\rm eex\bar x$, which is $\rm ex$, $\rm e\bar x$, $\rm eex$ or $\rm ee\bar x$,
is a tightly bound system which may be treated as a point particle.
(3)The particles $\rm x$ and $\rm \bar x$ are confined to each other within the distance of $\rm 1{\text -}3\, nm$,
which means the system $\rm eex\bar x$ as a whole is a sturdy system of the same scale.

Then we note that there are mobile states of $\rm eex\bar x$ in an HTS.
To proceed further we need an extra assumption:
\begin{quote}
{\bf Assumption E:} For a wave function of $\rm eex\bar x$ which is mobile in a solid,
the density of a negatively charged subsystem of $\rm eex\bar x$ anywhere in the inner space
of the atoms and the ions which constitute the solid
cannot be larger than its mean density in the outer space.
\end{quote}
Under the assumption a lower bound for the energies of mobile states can be obtained from a model
in which the negative charges are evenly distributed throughout the solid.

The calculation depends on the actual forms of Coulomb potential inside the atoms and the ions
which constitute the solid.
In particular the ratio of the inner space volume to the total volume is crucial
since a higher ratio means that the negative charges stay more in the inner space than in the outer space.
Since our goal is to obtain a rough lower bound for the mobile state energies
we consider a model specified by the following four conditions:
(1)The lattice has the simple cubic structure whose edge is $\rm 0.3\, nm$ wide.
(2)An ion with charge $e$ is located at every vertex of the lattice and the radius of the ion is $\rm 0.15\, nm$.
(3)In the inner space the potential of an electron at distance $r$ from the nucleus is
$-\frac{\epsilon}{r^2}$ with $\rm \epsilon=0.22\, eV\cdot nm^2$.
(4)The outer space potential for an electron is homogeneously $\rm -9.6\, eV$.

Note that $-\frac{\varepsilon}{r}$ is the potential of an electron at distance $r$ from a proton,
where $\varepsilon = \rm 1.22\, eV\cdot nm$.
In fact the constant $\epsilon$  in (3) above is chosen
so that $-\frac{\epsilon}{r^2}=-\frac{\varepsilon}{r}$  when $r=\rm 0.15\, nm$.
If $Z$ is the atomic number of the ion,
the inequality $-\frac{Z\varepsilon}{r} \leq -\frac{\epsilon}{r^2}\leq -\frac{\varepsilon}{r}$ holds
when $\frac{a}{Z} \leq r \leq a$, where $a=\rm 0.15\, nm$.
Thus $-\frac{\epsilon}{r^2}$ is a reasonable choice at least in the interval $\frac{a}{Z} \leq r \leq a$.
Furthermore since our goal is to find a rough lower bound our choice in (3)
can be justified in the whole interval $0< r \leq a$.
The homogeneous potential $\rm -9.6\, eV$ for the outer space in (4) is chosen
considering the facts: (i)We have that $-\frac{\epsilon}{a^2}=\rm -9.6 eV$.
(ii)There are free electrons present in the outer space.
(iii)$\rm -9.6\, eV$ is a value close to the minus of the sum of the work function $\rm 4\, eV$
and the Fermi energy $\rm 4\, eV$ of a typical metal.

Let us consider firstly the case when the structure is $\rm ex{\text -}e\bar x$.
Then we obtain $\rm -31\, eV$ for the contribution of the inner space
regardless of the value of $q_x$, $0 \leq q_x \leq e$.
Then by adding the contribution of the outer space we obtain $\rm -40\, eV$
as a lower bound of the mobile state energy of $\rm ex{\text -}e\bar x$.
Now we consider the work function of a typical metal and take $\rm -4\, eV$
as the energy of the electrons at the highest state.
Then the inequality $E_x < E_0$ means that an upper bound for the mass excess
of $\rm ex{\text -}e\bar x$ is $\rm 32\, eV$.
When the structure is $\rm x{\text -}ee\bar x$ and $q_x=e$, a similar calculation implies
that the upper bound is $1.5$ times $\rm 40\, eV$ minus $\rm 8\, eV$, that is, $\rm 52\, eV$.
Note that $\rm 32\, eV$ and $\rm 52\, eV$  are respectively the smallest and the largest values
of all the upper bounds calculated as in the above for all the possible structures of $\rm eex\bar x$
and for all the values of $q_x$, $0 \leq q_x \leq e$.

The value $\rm -40\, eV$ in the above cannot be too far above a value
obtained by more faithfully reflecting the lattice structure of a real HTS
and adopting more realistic potentials, as long as the calculation is based
on the uniform distribution of the negative charges.
Rather the fact might be more serious that, as a large system
and also due to the unknown interaction between $\rm x$ and $\rm \bar x$,
$\rm eex\bar x$ may remain mobile
while the negatively charged subsystems stay more in the inner space than in the outer space.
Nevertheless the author believes Assumption E cannot be violated too much
and the upper bound of the mass excess is most likely less than $\rm 100\, eV$.
To be safer, the author proposes that the mass excess of $\rm eex\bar x$  is less than $\rm 200\, eV$.
He admits that it is very unlikely that one may provide a low and solid enough upper bound
for the mass excess without a proper understanding of the interactions.

If the mass excess of $\rm eex\bar x$ were too small, HTS materials would have been much more common.
The author proposes that the mass excess is above the typical Fermi energy of a metal,
which is $\rm 4\, eV$, and, as stated in the above, below $\rm 200\, eV$.

\section{Low energy electron-electron scattering} 
The basic assumption is that the bound system $\rm eex\bar x$ may exist in free space
and its lifetime is short enough. Then in an electron-electron beam scattering arrangement
we may expect that there is a resonance for the formation of $\rm eex\bar x$
when the energy is the same as the mass excess of $\rm eex\bar x$ relative to two free electrons.
The kinetic energy of the electrons should be between $\rm 2\, eV$ and $\rm 100\, eV$
if we take seriously the proposal of the previous section. The formation of $\rm eex\bar x$ will be shortly
followed by a decay into two electrons. We do not know whether the decay will accompany emission of photons.
Therefore we propose that the following is what we should search for:
\begin{quote}
A peak of events, where each event is such that two electrons are simultaneously
scattered off from the same location in directions perpendicular to the beams.
\end{quote}
Then another basic assumption is that the cross section of the event, $\rm e+e \rightarrow eex\bar x$,
at the peak is large enough to be detected by the experiment.

For the two electrons originated from a decay of $\rm eex\bar x$,
a direction perpendicular to the beams is not special in any sense.
The plane perpendicular to the beams is as likely as any other planes through the location of $\rm eex\bar x$
for both electrons to travel on.
On the other hand assume two electrons are directed to each other with the same speed in opposite directions.
When there is no intermediate bound system it cannot be expected that both are scattered
in directions perpendicular to their initial directions, regardless the scattering is elastic or not:
Assume on the contrary that both electrons were scattered in perpendicular directions.
Then one may not tell which of the scattered electrons is more likely the electron, say,
from the left rather than from the right since they are identical and the outcomes are symmetric.
Furthermore since they are Fermions, any contribution to the cross section
calculated assuming a specific track cancels out with the other contribution.
Thus in principle there is no background noise coming from a perpendicular scattering of both electrons by the usual process.

However an electron may come out in a perpendicular direction by a usual inelastic scattering
while the other electron moves in a non-perpendicular direction.
And there is a chance that two independent such events occur within the time resolution of the detector
at some indistinguishably close locations.
This noise can be made arbitrarily small relative to the rate by which the target events may occur
either by enhancing the time resolution or by decreasing the event rate,
that is, by decreasing the luminosity of the beams:
First of all we note that in reality we have to detect electrons scattered off in directions
which make less than a fixed angle with the perpendicular plane.
Let us refer to these directions as {\it nearly perpendicular} or simply as perpendicular
if there is no possibility of confusion.
Let $\rho$ denote the rate per unit time by which a perpendicular scattering of an electron by a usual process may occur
at a location within a small region $R$ whose two points are indistinguishable.
Note that $\rho$ is proportional to the luminosity of the beams. Let $\Delta \tau$ denote the time resolution of the detector.
Then $\rho^2 \Delta \tau$ is the rate per unit time by which false simultaneous perpendicular scattering events may occur.
However $\kappa\rho$ is the rate per unit time by which $\rm eex\bar x$'s may be formed in $R$
and both of the electrons are scattered off in perpendicular directions, where $\kappa$ is a non-negative real number
which is unrelated to $\rho$ and $\Delta\tau$ and supposedly depends on the energy of the electron beams.

Let us assume that a peak can be recognized as such only when $\rho^2\Delta \tau \leq \kappa\rho$, that is,
when $\rho\Delta\tau \leq \kappa$.
Note that $\kappa$  is in fact the cross section of the event, $\rm e+e \rightarrow eex\bar x$,
multiplied by the ratio of the near perpendicular planes through $R$ to all the planes through $R$
and then divided by the cross section of the event
in which an electron is scattered off in a perpendicular direction by usual scattering.
Therefore it is desirable that $\rho\Delta\tau$ should be kept as small as possible
so that a peak can be recognized even when the cross section of $\rm e+e \rightarrow eex\bar x$ is very small.
On the other hand, since the real detector will register electrons which are scattered in not exactly but nearly perpendicular directions,
there should be some noise $\nu\rho$
which is the rate by which a nearly perpendicular scattering of both electrons may occur by the usual process,
where $\nu$ is a small positive real number. Note that it is unnecessary for one to keep $\rho^2\Delta\tau$ much less than $\nu\rho$.
For simplicity we assume that one arranges the experiment so that $\rho\Delta\tau =\nu$.
This means that the peak can be detected only when $\nu\leq \kappa$, which is reasonable.
Thus one should keep $\rho\Delta\tau$ and $\nu$ as small as possible.
However $\kappa\rho$ should be reasonably large so that the target events may occur
at a rate high enough not to hinder the experiment.
On the other hand $\nu$ depends on the largest angle which a nearly perpendicular plane makes with the exactly perpendicular plane.
Thus a smaller $\nu$  means a smaller $\rho$ if the same luminosity is maintained.
This means that one needs an intensive beam to keep $\nu$ small while maintaining $\rho$ large enough.
Therefore both the intensity of the beam and the time resolution of the detector are important.

\vspace{11pt}
\noindent {\bf Concluding remarks:} Suppose one has discovered, by chance for instance,
a resonance as described in the above without any reference to this paper.
Then still it should have meant an unstable bound state of two electrons
with a mass excess corresponding to the peak energy.
However it would not have directly implied the $\rm eex\bar x$-hypothesis.
The same holds even if a resonance happen to be found exist by an experiment influenced by this paper.
Therefore the paper could have been organized without any reference to the $\rm x$-particle as follows:
1)Propose that there is a bound state of two electrons with a mass excess.
2)Point out that, if there is a low enough state for the bound electrons in a solid,
the bound system in the state will be stable.
3)Discuss as in \S 2 above the properties of a solid which admits such a state.
4)Estimate as in \S 4 the mass excess. 5)Propose an experiment as in \S 5.
Most of the discussions on the interaction of $\rm x$ and on the binding mechanism in \S 3
above could have been omitted in the other version.
However it would have been truly difficult for the author, for a reader or
for any one else who proposes a binding of two electrons with a mass excess
to refrain from the question of what is the binding mechanism.

\end{document}